\documentclass[a4paper, USenglish]{lipicsoid-2019}

\hideLIPIcs
\usepackage{etex}
\usepackage{amsmath}
\usepackage{amssymb,amsthm,url}
\usepackage{stmaryrd}
\usepackage{todonotes}
\usepackage{comment}
\usepackage[numbers, sort&compress]{natbib} % improved bibliography
\usepackage{doi} % hyperlink doi numbers
\usepackage{bibentry}
\nobibliography*

\hypersetup{
  colorlinks=true,
  linkcolor=black,
  citecolor=black,
  filecolor=black,
  urlcolor=[rgb]{0,0.1,0.5},
  pdftitle={Note on distributed certification of minimum spanning trees},
  pdfauthor={}
}

\title{Note on distributed certification\newline of minimum spanning trees}

\author{Laurent Feuilloley}%
{Sorbonne Université and Universidad de Chile}%
{feuilloley@dii.uchile.cl}%
{http://orcid.org/0000-0002-3994-0898}%
{}

\authorrunning{L. Feuilloley}
\funding{Support by ANR ESTATE}

\begin{document}
\maketitle

%%%%%%%%%%%%%%%%%%%%%%%%%%%%%%%%%%%%%%%%%%%%%%%%%
%%\vspace{1cm}
%%%%%%%%%%%%%%%%%%%%%%%%%%%%%%%%%%%%%%%%%%%%%%%%%

\begin{abstract}
A distributed proof (also known as local certification, or proof-labeling scheme) is a mechanism to certify that the solution to a graph problem is correct. 
It takes the form of an assignment of labels to the nodes, that can be checked locally.
There exists such a proof for the minimum spanning tree problem, using $O(\log n \log W)$ bit labels (where $n$ is the number of nodes in the graph, and~$W$~is the largest weight of an edge). 
This is due to Korman and Kutten who describe it in concise and formal manner in~\cite{KormanK07}. 
In this note, we propose a more intuitive description of the result, as well as a gentle introduction to the problem.\\

\noindent \emph{This note originates from a careful reading of \cite{KormanK07}, while working on~\cite{BlinDF19}. Comments are most welcome.}
\end{abstract}

%%%%%%%%%%%%%%%%%%%%%%%%%%%%%%%%%%%%%%%%%%%

\section{Introduction}
\label{sec:introduction}

\paragraph*{Distributed checking of minimum spanning tree.}

The problem is the following: we are given a weighted graph where some edges are selected, and we want to enable the nodes of the graph to collectively check whether the selected edges form a minimum spanning tree (MST) of this weighted graph or not.

More precisely, we want the nodes to take a \emph{distributed decision}. For this, every node takes its own decision whether to accept or to reject, and the collective decision is considered to be an acceptance, if and only if, all nodes accept.
The difficulty is that the nodes have a limited knowledge of the graph.
Each node has a local view that contains only its adjacent edges and nodes, along with the weight of these adjacent edges and whether they are selected or not.
In addition, we assume that the nodes are given unique identifiers.

Actually we will need to provide more information to the nodes. Indeed, it is impossible to check whether the selected edges form a minimum spanning tree or not in this restricted setting, as the following example shows. 
Consider a ring, where the nodes have arbitrary distinct identifiers, and all edges are selected.
Now take an arbitrary node. 
Given its local view, it cannot distinguish whether it is indeed in a ring, or if it is in the middle of a long path, where all edges are selected. 
If this node chooses to reject, then the distributed decision will automatically be a rejection, and in the long path this would be a wrong decision. 
Thus it has to accept. 
But then, in the ring, by symmetry, every node has to accept, and then the distributed decision is an acceptance, although the ring is not a correct instance.

\paragraph*{Distributed proof.}

The mechanism used to bypass the impossibility above is called a \emph{distributed proof}. 
In such a mechanism, every node will be given a label, and a node can see not only its own label but also the ones of its neighbors.
These labels are supposed to certify the correctness of the minimum spanning tree, in the following sense: 
\begin{itemize}
\item If the set of selected edges form a minimum spanning tree, then there exists a label assignment such that all nodes accept.
\item If the set of selected edges does not form a minimum spanning tree, then for any label assignment, at least one node will reject. 
\end{itemize}
In other words, the set of labels certifies that the configuration is correct, and the nodes can check this certification.

\paragraph*{A bit of vocabulary and notations.}

Let us introduce some terminology. 
The set of selected edges is \emph{the solution}, and the graph and the solution form  together \emph{the configuration}.

The labels used in a distributed proof are called the \emph{certificates}, and a distributed proof may also be called a \emph{distributed certification}. (The original name for distributed proof is \emph{proof-labeling scheme}.)
When describing the mechanism, it is convenient to refer to an entity that assigns the labels to the nodes, this is \emph{the prover}. 
Note that we are not interested in the question of how the prover computes the certificates. 
Also note that given the definition of certification, we can see the prover as an entity trying to make the nodes accept, regardless of whether the solution is correct or not. And the nodes are capable of detecting when the prover is trying to fool them. When describing a scheme, we only describe what the prover does on correct instances, as on incorrect instances it can assign arbitrary labels.

The number of nodes in the graph is denoted by $n$, the weight of an edge $(u,v)$ is denoted by $w(u,v)$, and the maximum weight is $W$. 
We assume that the identifiers are on $O(\log n)$ bits.
The size of a distributed proof for MST, is a function of $n$ and $W$, and is the size of the largest label for an instance of $n$ nodes, and max weight $W$. 

We refer to the survey~\cite{FeuilloleyF16} for more information about distributed decision and certification.

\section{Warm-up: non-optimal MST distributed proofs}
\label{sec:non-optimal}
The goal of this note is to explain the MST distributed proof of~\cite{KormanK07}, which uses $O(\log n  \log W)$ bits. 
This size is optimal, as proved in~\cite{KormanK07}. 
As a warm-up, we sketch a few simpler distributed proofs for minimum spanning tree, that use larger certificates. 

\paragraph*{Universal proof.}
A basic result about distributed certification is that it is possible to certify any solution of problem~\cite{KormanKP10}. 
On a correct instance, the prover just has to provide to every node a complete description of the configuration. 
More precisely, every node is assigned a certificate containing: (1) the adjacency matrix of the graph, (2) the bijection saying which row of the matrix corresponds to which node of the graph (designated by its identifier) and (3) a description of the solution (and extra input information when needed, \emph{e.g.} edge weights for minimum spanning tree). 
It is then easy for the nodes to check this distributed proof: every node checks that it has the same certificate as its neighbors, that its local view is consistent with the graph described in the certificate, and that the solution is correct in this described graph.
The problem of this technique is that it requires large certificates: the adjacency matrix alone takes $\Omega(n^2)$ bits.

\paragraph*{Proofs based on algorithms and `à la Kruskal' proof.}

A classic technique in distributed certification is to take a distributed algorithm that can produce the solution at hand, and to describe its run in the certificates. 
More precisely, the prover can give to every node the list of all the messages that it would send and receive during the run. 
Then the nodes can check that this run is correct, by virtually running the algorithm, based on this transcript, and they can check that it produces the solution at hand. 
The certificates given by this technique are unfortunately very large in general.
A strategy is to keep only the essential pieces of information, and to design to make the node not verify the run {\emph{per se}}, but only some key properties of it. 

Let us exemplify this technique with Kruskal's algorithm. 
In this algorithm (which is a centralized algorithm), the edges are ranked by increasing weight, and are considered in this order. 
Each edge is selected to be in the minimum spanning tree if it does not create a cycle with the edges already selected.
The certificates that are derived from this algorithm are the following. 
Every node is given the list of the selected edges (described by the identifiers of the two endpoints), along with the weights of these edges.
It is proved in \cite{FeuilloleyH18} that this is enough to allow the nodes to check the solution using a procedure based on Kruskal algorithm: checking at every step that no lighter edge could be added without closing a cycle.
This scheme uses labels of size $\Theta(n \log n + n \log W)$, which is better than the universal proof but still very large.

\paragraph*{`À la Borůvka' proof.}

Yet another scheme is based on another MST algorithm, namely Borůvka's algorithm (or its distributed variant the GHS algorithm~\cite{GallagerHS83}).
It uses $O(\log^2 n +\log n\log W)$ bit certificates~\cite{KormanKP10}. 
The lower bound for MST certification being $\Omega(\log n \log W)$~\cite{KormanK07},  this scheme is optimal for the regime where $W$ is at least polynomial is $n$. 

Let us remind Borůvka's algorithm. 
It proceeds in phases, using objects called \emph{fragments}, that are acyclic subgraphs. 
At the beginning of the run, every node is a fragment, and during the run of the algorithm, edges are added between fragment such that they get merged. 
At the end there is only one fragment, and it is a minimum spanning tree. 
The way the edges are added is the following.
There are phases, and at each phase every fragment chooses a so called \emph{outgoing edge} (that is an edge that has exactly one endpoint in the fragment) of minimum weight. 
These outgoing edges are then added to the solutions, and we proceed for the next phase.
At each phase a fragment merges with at least one other fragment, therefore there are at most $O(\log n)$ phases. 

In the certificates a node is given, for each of the $O(\log n)$ phases, the name and weight of the outgoing edge chosen by its fragment (along with a proof that this edge exists). 
This is enough to check that that the solution is an MST. 
It uses $\Theta(\log n +\log W)$ bits per phase, hence the size of the scheme is $\Theta(\log^2 n +\log n\log W)$.

\section{Theorem and outline}

The focus of this note is the following theorem.

\begin{theorem}[\cite{KormanK07}]
There exists a distributed proof for MST of size $O(\log n \log W)$. 
\end{theorem} 

\noindent The rest of the note is in an explanation of the proof of this theorem. 

\section{A basic tool: spanning trees}
\label{sec:tree}

\paragraph*{Certifying a spanning tree.}
Before certifying a minimum spanning tree, it is useful to know how to certify a spanning tree. 
The classic scheme is the following: on correct instances, the prover chooses an arbitrary node as the root of the tree, and assigns to every node $u$ a certificate containing the identifier of the root, and the distance from $u$ to the root.
It is folklore that this is enough to certify a spanning tree: the acyclicity is certified by the distance, and the connectivity is certified by the root identifier (see \cite{FeuilloleyF16} for explanations).
The two pieces of information above (the identifier and the distance) both use $\Theta(\log n)$ bits, thus certifying a spanning tree can be done using $\Theta(\log n)$ bits. 

\paragraph*{Uses of a certified spanning tree.}

The first use we make of the certification scheme above is to simply certify that the solution at hand is a spanning tree. 
This way, we can focus on proving the minimality. 
The solution can now be called \emph{the input tree}.

Second, we will use other spanning trees than the input tree. 
More precisely, in order to certify the minimality, the certificates will contain a distributed description and the certification of spanning trees different from the input tree. 
For example, suppose that at some point you want to certify that there is a node with identifier $x$ in the graph. 
To do so, the prover can describe and certify a spanning tree of the graph whose root is this special node. 
As this spanning tree is certified, the nodes will be able to check its structure, and if the root also checks that is has identifier $x$, and no node rejects, it means that indeed a node with identifier $x$ exists.

\section{Idea 1: Checking the cycle property}
\label{subsec:checking-weights}

Now that we have the spanning tree tool, let us start to discuss the ideas that lead to the distributed proof of~\cite{KormanK07}.
The first idea is to differ from the approaches described in Section~\ref{sec:non-optimal}, by \emph{not} mimicking an algorithm. 
Instead, we will allow the nodes to check a fundamental property of minimum spanning trees: the \emph{cycle property}.

\paragraph*{Cycle property.}
Consider a spanning tree $T$ of a weighted graph. 
For any edge $(u,v)$ of the graph, let $\max_T(u,v)$ be the largest weight on the path from $u$ to $v$ in $T$. 
The cycle property states that $T$ is minimum, if and only if, for every edge $(u,v)$,  $w(u,v)\geq \max_T(u,v)$. 
Hence if the nodes get to know the values $\max_T(u,v)$, they can check the inequality for all edges, and thus check the minimality.

\paragraph*{Naive approach.}
A naive idea consists in writing in the certificate of every node $u$, the values $\max_T(u,v)$, for all neighbor $v$ of $u$.
Without even looking into how the nodes could check that these values are correct, there is already a problem of size: this can take $\Theta(n\log W)$ bits per node, as we have no control on the degree.
Thus, we want to allow every node $u$ to retrieve the values $\max_T(u,v)$, without explicitly writing them in the certificate. 

\section{Idea 2: Using intermediate nodes}

\paragraph*{Using non-neighbor intermediate nodes}
In the naive approach above, a node would not use the fact that it can see the certificates of its neighbors. 
But if $v$ is a neighbor of $u$, then $u$ has access to the certificate of $v$.
Now, an important idea, is that it can be useful for a node $u$, to know the value $\max_T(u,w)$, even if the node $w$ is not a neighbor of $u$.
Indeed, if $w$ is a node on the path between $u$ and $v$ in the input tree, and if $u$ and $v$ respectively hold $\max_T(u,w)$ and $\max_T(v,w)$, then they can compute $\max_T(u,v)$, because:
\[
{\max}_T(u,v)=\max({\max}_T(u,w),{\max}_T(v,w)). 
\]
In this case, it is useless that either $u$ or $v$ stores $\max_T(u,v)$ as they can compute it.
Now our scheme will be based on the idea of giving to any node $u$, a list of values $\max_T(u,w)$, for some carefully chosen $w$.
For a given node $u$, such an intermediate node $w$ will be called \emph{an anchor} of $u$.
Note that we actually need to provides pairs $(ID(w),\max_T(u,w))$, where $ID(w)$ is the identifier of $w$, in order to allow the nodes to compare their sets of anchors.

\paragraph*{Which nodes to use?}

Now the question is: how to chose the sets of anchors?
This set of anchors has to fulfill several goal. 
First, it should allow any pair of neighbors $u,v$ to retrieve the value $\max_T(u,v)$, thus any pair of neighbors should have an anchor in common, and this anchor should be on the path between them in the input tree. 
Second, we want the nodes to be able to check that the pairs $(ID(w),\max_T(u,w))$ are correct.
Third, we want to minimize the number of pairs $(ID(w),\max_T(u,w))$ that any node gets.

\section{Idea 3: Ancestors as anchors}

Remember that thanks to Section~\ref{sec:tree}, we can assume that the solution is a spanning tree of the graph, and that it is oriented towards a root.
We will use this structure to chose our anchors.

\paragraph*{Ancestors.}

We simply use the ancestors of $u$ in the tree as its anchors (including $u$ itself). 
We claim that this is a correct set of anchors.
Indeed the nearest common ancestor of $u$ and $v$ is an anchor of both $u$ and $v$ as it is an ancestor of both, and it is on the path between the two nodes.
But the nodes may have many ancestors in common, and only the nearest common ancestor is on the path between them. How to allow the nodes to identify the nearest common ancestor? 
We simply list the pairs $(ID(w),\max_T(u,w))$, from the closest ancestor to the root. This way the nodes simply have to look for the first anchor that is on both lists.

\paragraph*{Checking.}
Now we need to describe how the nodes check that the values they are given are correct. 
For the ancestors, every node just has to check that its ancestors list is consistent with the lists of its parent and children in the tree. 
For the maximum weight, similar checking works, except that now the nodes also check that the weights of their local views are consistent with the certificates.   

\paragraph*{Size.}

At this point, we have described a correct distributed proof for minimum spanning tree, based on checking the cycle property. A problem is again the size.
If $d$ is the depth of the tree with the orientation we took at the beginning the scheme described above has size $\Theta(d\log n+d\log W)$. 
As we can control the diameter of the input tree, the size can be as large as $\Omega(n\log n + n\log W))$, for example when the tree is a path.
 
\section{Idea 4: Using another tree}

An idea to cope with the depth problem is to define the anchors using a tree that is \emph{not} the input tree.
That is, the anchors of a node will still be its ancestor in a tree, but not the same tree. 
This new tree will be called \emph{the overlay tree}.
We will make sure that this overlay tree is balanced, in order to get a small depth, and avoid the problem of the previous section.

\paragraph*{Virtual edges}

In general, the overlay tree cannot be defined as a spanning tree of the graph. Indeed, if the graph is a path, for example, it has no small-depth spanning tree. 
Instead, we will use virtual edges, that is the overlay tree may have some edges between nodes $a$ and $b$, even if there is no edge $(a,b)$ in the graph. 
Of course this has to be done carefully, and will make the certificates more complicated. 

\paragraph*{Top-down construction of an overlay tree}

Let us now describe a general construction to build an overlay tree. 
We will take care of the depth later. 
The overlay tree is built in a top-down fashion. 
Take an arbitrary node, and make it the root of the overlay tree. 
After removing it from the input tree, we are left with some $k$ connected components (of the input tree).
For each of them chose an arbitrary node, these are the $k$ children of the root. 
Continue recursively until all the nodes appear in the overlay tree.

\paragraph*{The nearest common ancestor is still a correct anchor choice}

We claim that the construction above preserves the following key property: given two nodes $u$ and $v$, their the nearest common ancestor \emph{in the overlay tree}, is on the path between them \emph{in the input tree}.
At each step of the top-down construction of the tree, as long as $u$ and $v$ remain in the same connected component, they are assigned the same ancestors.
Then at some point a new node~$c$ is chosen to be the root of one of the connected components, and its removals leaves $u$ and $v$ in different components. 
On the one hand, as the removal of $c$ separates $u$ and $v$, the node $c$ must be on the path from $u$ to $v$ in the input tree. 
On the other hand, $c$ is the nearest common ancestor they have in the overlay tree. 
This proves the claim.

\paragraph*{Encoding and checking of the overlay tree.}

The certificate of a node contains, as before, a list of pairs $(ID(w),\max_T(u,w))$, 
but the checking that the nodes $w$ are ancestors in the overlay tree requires more work than the analogous checking in the original tree. 
Let $w$ be one of these ancestor nodes. How can the node $u$ check that $w$ really exists and is an ancestor? Or more precisely, how can we encode the overlay tree, to make this checking possible. 

By construction, the nodes that have $w$ as an ancestor in the overlay tree form a connected component in the input tree. 
We can then define a tree that is spanning this component, and whose root is $w$. 
Then, the certificate of $u$ will contain, in addition of $(ID(w),\max_T(u,w))$, the local encoding and certification of this tree. 
That is, it contains the name of the parent of~$u$ in this `component tree' as well as the distance from $u$ to $w$ in this tree.
As this component tree is based on edges of the graph (contrary to the overlay tree), it can be checked easily in the same way as in Section~\ref{sec:tree}.

We do so for every ancestor of every node. That is, a node is given the local encoding and certification of a spanning tree for each of its $O(\log n)$ ancestors.
Not only does this ensures the ancestors really exist, but by checking the consistency of these spanning trees, the nodes can also check the whole overlay tree structure.  

\paragraph*{Balanced overlay tree.}

In the construction of the overlay tree, there is freedom for the choice of the root. We can chose this node to be a center of the tree, that is a node such that its removal leaves the tree with components of size at most $n/2$. 
By choosing such a center at each step of the recursion, we get an overlay tree that is balanced, and has depth in $O(\log n)$.

\paragraph*{Size.}
Let us now consider the size of this new scheme. 
We still have, in the first part of the certificate, the certification that the input tree is a spanning tree of the graph, which takes $\Theta(\log n)$ bits. 
This is negligible compared to what follows (and it will also be in the next sections).
In addition to this, the certificate of a node $u$ contains $O(\log n)$ fields (one for each ancestor in the overlay tree), with:
\begin{itemize}
\item the identifier of the ancestor in the overlay tree,
\item the maximum weight on the path between $u$ and the ancestor in the input tree,
\item the identifier of a neighbor of $u$ (tha parent in the component tree), 
\item a distance (from $u$ to its ancestor through the component tree).
\end{itemize} 

In total, the size is in $\Theta(\log^2\!n+\log n\log W)$. 
Therefore, this scheme is as good as the scheme based on Borůvka algorithm, but not better. 
In particular, even if $W$ is very small, the size will still be in $\Omega(\log^2\!n)$.
In the next sections, we show how to make the $\log^2\!n$ term disappear, to get the $\Theta(\log n\log W)$ size. 
To do so, we first encode the overlay tree in a more compact manner (to avoid using the identifier of a neighbor at each level), and then make the names of ancestors more compact too (to avoid using the identifier of an ancestor at each level).

\section{Idea 5: Compact encoding of the overlay tree}

To describe and check the structure of the overlay tree, we can do better than giving the identifier of a neighbor for each of the $O(\log n)$ levels. 
To do so, we will rely on the input tree (and its orientation to a root) that we have. 
 
For each level of the overlay tree, instead of giving the identifier of a neighbor as a pointer to the parent in the component tree, we simply write in which direction this neighbor is with respect to the input tree. 
That is, if this neighbor is the parent in the input tree, then the label will simply say~\emph{up}, otherwise, it is a children in the input tree, and the label says \emph{down}. 
(The third case being that if the node is the root of this tree structure, then it has a label \emph{root}.)

Two things to point out here. First, as the input tree is certified to be acyclic, there is no risk to have something like a cycle of nodes with label \emph{up}.
Second, if the flag is \emph{down}, at first a node cannot know which of its children is the good one. But actually, if the labeling is correct, exactly one of its children has a label that is not \emph{up} (that is either \emph{root} or \emph{down}), and this is its parent.
Thus, from this up-down labeling, the nodes can recompute the identifier of their parents, and check the correctness.
The good things is: this takes only $O(1)$ bits. 
Thus in total this is $O(\log n)$ bits, hence we have eliminated the first $\log^2n$ term of the size.

\section{Idea 6: Compact encoding of the ancestors}

The remaining $\log^2\!n$ term comes from the list of $\Theta(\log n)$ ancestor identifiers that each node holds. 
Note that it is crucial to have some way of naming these ancestors, such that two neighbors of the graph can decide which anchor of their lists they should use to compute the max value, and then check the cycle property. 
Also, in general, there is no better way of naming a set of $\Theta(\log n)$ nodes than to use $\Theta(\log^2n)$ bits. 
But, the sets of nodes we are interested in are not arbitrary: they are chains of nodes that form a path from a node to the root, through all its ancestors.
We can tweak the names such that retrieving the concatenation of names of such chains requires small labels. 

\paragraph*{Subtree numbering to avoid identifiers}

A way to give distinct names to the node of a tree is the following: 
every node gives a distinct number to each of its $k$ children (from 1 to $k$), in an arbitrary way. 
Then the name of a node is the concatenation of the numbers of all its ancestors from the root to itself. 
It is easy to check that these names are unique. 

Now, in our context, this is especially useful, because it is not necessary anymore to give the name each ancestor: from its own name, a node can retrieve the names of all its ancestors, as they are just prefixes.
Also, because of the overlay tree, the nodes can check that these new names are correct.

We are not yet done, as the new names might have $\Omega(\log^2n)$ bits, for example if a node has all its ancestors having (almost) $\Omega(n)$ children.  

\paragraph*{Compact subtree numbers}
We can play with the way a node numbers its children.
Namely, we number the subtrees in a similar fashion as in \cite{GavoilleKKPP01}, that is, in the inverse order of the size of the subtree. 
More precisely, we number the subtrees from~1, for the subtree with the largest number of nodes, to $k$, for the subtree with the smallest number of nodes. 

Let us compute the size of the name of an arbitrary node $u$ at depth $r$.
For every $i$, from~0 to~$r$, let $a_i$ be the ancestor of $u$ of depth $i$ (with $u=a_r$). 
Also let $n_i$ be the number of nodes in the tree of $a_i$ (that is $a_i$ and all its descendants).
Finally, for $i\geq 1$, let $k_i$ be the number given to $a_i$ by its parent. 
Now, if a node is given a number $k_i>1$, it means that its parent had at least $k_i-1$ children with larger trees. 
Thus for all $i$ from 1 to $r$, $n_{i-1}\geq k_i\cdot n_{i}$, that is $k_i\leq n_{i-1}/n_{i}$.
Then the size of the name of $u$, wich is the size of the concatenation of the subtree numbers of the ancestors of $u$ is of order:
\[
\sum_{i=1}^{r} \log(k_i) 
= \log(\Pi_{i=1}^{r} k_i)
\leq \log \left( \Pi_{i=1}^{r} \frac{n_{i-1}}{n_{i}} \right)=\log\left(\frac{n_0}{n_r}\right) \leq \log (n_0)=\log (n).
\]

That is the name of every node is on $O(\log n)$ bits, thus we have eliminated the second $\log^2n$ term.\footnote{A subtlety is that, to be able to retrieve the subtree numbers from the concatenation, one has to add commas between the numbers, but as there are at most $O(\log n)$ ancestors, this incurs no overhead. Actually, more complex nearest-common ancestor labelings can cope with more ancestors (see\cite{AlstrupGKR}, used in \cite{BlinF15}).}

\section{Wrap-up}

Putting all the pieces together, we get $O(\log n \log W)$ bit labels for certifying MST. 
This is optimal~\cite{KormanK07}.
Something that may be a bit surprising at first, is that the nodes cannot check that the overlay tree is balanced and that the subtree numbering is an inverse-size numbering.
This is no problem for our setting, as we only want to (1) certify the solution, and (2) have small labels on correct instances. 
This can be problematic if we want to build the certificates in a distributed manner without exceeding some space limit. 
A recent preprint~\cite{BlinDF19} tackles the problem of how to construct the labels to a get an optimal space self-stabilizing algorithm, and solve this problem by allowing the nodes to check the size of their labels.

%%%%%%%%%%%%%%%%%%%%%% biblio %%%%%%%%%%%%%%%%%%%
% \def\UrlFont{\sf\footnotesize}
\DeclareUrlCommand{\Doi}{\urlstyle{same}}
\renewcommand{\doi}[1]{\href{https://doi.org/#1}{\footnotesize\sf doi:\Doi{#1}}}

\bibliographystyle{plainnat}
\bibliography{biblio-certification-MST}
%%%%%%%%%%%%%%%%%%%%%%%%%%%%%%%%%%%%%%%%%%%%%%%%

\end{document}